\documentclass[aps,prb,twocolumn,showpacs,floatfix]{revtex4}

\usepackage{graphicx}

\begin{document}

\title{Superconducting thin rings with finite
 penetration depth}

\author{Ernst Helmut Brandt}
\affiliation{Max-Planck-Institut f\"ur Metallforschung,
   D-70506 Stuttgart, Germany}
\author{John R.\ Clem}
\affiliation{Ames Laboratory - DOE and Department of Physics and
Astronomy, Iowa State University, Ames Iowa 50011, USA}

\date{\today}

\begin{abstract} Recently Babaei Brojeny and Clem [Phys.~Rev.~B {\bf
68}, 174514 (2003)] considered superconducting thin-film rings   in
perpendicular magnetic fields in the ideal Meissner state with
negligibly small magnetic penetration depth and presented useful
analytical limiting expressions and numerical results for the
magnetic-field and sheet-current profiles,
 trapped magnetic flux, self-inductance, magnetic moment, and
focusing of magnetic flux into the hole when no net current flows in
the ring. The present paper generalizes all these results to rings
with arbitrary values of the two-dimensional effective penetration
depth
$\Lambda = \lambda^2 /d$ ($\lambda$ is the London depth and
$d < \lambda/2$ the film thickness) using a straightforward matrix
inversion method. We also present results for the energy of a
superconducting ring as a function of the applied magnetic
induction $B_a$ and the quantum number $N$ defining the size of the
fluxoid $N \phi_0$ trapped in the hole.
 \end{abstract}


\maketitle

\section{Introduction}  

In a recent paper Babei Brojeny and Clem \cite{1} calculated the
magnetic properties of superconducting thin-film rings and disks in
the ideally screening Meissner state when a perpendicular magnetic
field is applied. They showed that the Biot-Savart law for the
sheet-current density
${\bf J}(r) = \int_{-d/2}^{d/2} {\bf j}(r,z) dz$ (${\bf j}$ is the
current density) in a circular disk of thickness $d$ and radius $b$
with a hole of radius $a$ ($0 \le a < b$) can be solved by a
physical ansatz containing a factor
$[(r^2 -a^2)(b^2-r^2)]^{-1/2}$ describing the divergence of $J(r)$
at the inner and outer radii. A plethora of useful numerical results
and analytical limiting expressions is given in Ref.~\onlinecite{1}
for several situations: (a) magnetic flux trapped in the hole when
no magnetic field is applied; (b) zero magnetic flux in the hole
when the ring is subjected to an applied field; and (c) focusing of
the magnetic flux into the hole when a magnetic field is applied but
no net current flows around the ring.

  Throughout the paper \cite{1} it was assumed that either the
London magnetic penetration depth obeys $\lambda < d/2$ or, if
$\lambda > d/2$, the two-dimensional (2D) penetration depth
$\Lambda = \lambda^2 / d$ (or screening length \cite {1}
$2 \lambda^2 /d$) is negligibly small. The same assumption was made
in previous work on disks \cite{2,3,4} and rings.\cite{5,6} However,
while thin superconducting rings with
$\lambda =0$ ideally screen magnetic flux from penetrating the hole,
a finite $\lambda$ or $\Lambda = \lambda^2 /d > \lambda$ will allow
magnetic flux to penetrate into the film  as well as the hole. This
effect is much stronger than would be suggested by the exponential
factor $\exp[ -(b-a)/\lambda]$ that applies to long tubes in an
axial field.

  The effect of finite $\Lambda$ is particularly important for the
interpretation of experiments that try to confirm the cosmological
Kibble-Zurek mechanism \cite{7,8} of spontaneous formation of
vortices during rapid cooling of a superfluid; some of these
experiments use superconducting rings,\cite{9} while others use
disks.\cite{10} Finite $\lambda$ also modifies flux focusing,  an
important feature of SQUIDs (superconducting quantum interference
devices), which usually have roughly the shape of a washer with a
small central hole but with a slit that allows magnetic flux to
enter the hole and causes zero net circulating current at the SQUID's
critical current. \cite{11,12,Barone,Koelle} The concentration of
magnetic flux into this hole increases the effective area of the SQUID.

  This paper is organized  as follows. In Sec.~II we describe our
calculation method, which applies to arbitrary $\Lambda$.
In Sec.~III we compute the self-inductance of a thin flat ring.
In Sec.~IV we calculate the response of a ring in an applied
magnetic field. In Sec.~V we study the flux-focusing problem
and calculate the effective area.  In Sec.~VI we calculate the
energy of a ring as a function of the applied field and the quantum
number
$N$ describing the size of the fluxoid $N \phi_0$ trapped in the
ring.  In  Sec.~VII we give some analytical results for the limit
of large
$\Lambda \gg b$, which applies to mesoscopic rings.
We present a brief summary in Sec.\ VIII.

\section{Calculation method}  

We assume for simplicity
that current in a circular coil far from the ring
produces a vector potential ${\bf A}_a({\bf r}) = A_a({\bf r})
\hat\varphi$, which describes the magnetic induction
${\bf B}_a = \nabla \times {\bf A}_a$;  $\hat\varphi$ is the
azimuthal unit vector. Near the film in the plane $z = 0$,
we assume $A_a(r) = rB_a/2$,
such that the magnetic induction applied to the
thin-film ring is ${\bf B}_a = B_a \hat z$.
In response to either the applied field or a fluxoid trapped
in the hole, currents are induced in the ring.
The net magnetic induction is ${\bf B}({\bf r})
= {\bf B}_a({\bf r}) + {\bf B}_J({\bf r})$, where
${\bf B}_J({\bf r}) = \nabla \times {\bf A}_J({\bf r})$ and
its vector potential ${\bf A}_J({\bf r})$ are generated by the
currents in the film.  Because of the circular symmetry,
the sheet current in the film has only a
$\varphi$-component, ${\bf J}(x,y)=J(r) \hat\varphi$.
Similarly, the vector potential ${\bf A}(x,y,z)$, defining
the total magnetic induction ${\bf B} = \nabla \times {\bf A}$,
has only a $\varphi$-component. In the film plane $z = 0$,
we have ${\bf A}(x,y,0) = A(r) \hat\varphi$ and
$A(r) = A_a(r) + A_J(r)$. The current density ${\bf j}$ is
related to ${\bf A}$ via the London equation,
${\bf j} = -{\bf A}_s/\mu_0\lambda^2,$ where
${\bf A}_s$ is the superfluid velocity expressed in units
of vector potential; here ${\bf A}_s = A_s \hat\varphi$ with
  \begin{equation} 
  A_s(r) = A(r) -\Phi_f/2\pi r.
  \end{equation}
The second term on the right-hand side is due to the gradient
of the phase of the complex superconducting order
parameter.  For a ring with a slit we treat
$\Phi_f$ as a free parameter to be determined by boundary
conditions, but for an unslitted ring $\Phi_f$ corresponds to the
London fluxoid,\cite{13} which is quantized; i.e.,
$\Phi_f = N \phi_0$, where $N$ is an integer and
$\phi_0 = h/2e$ is the superconducting flux quantum.

It is useful to divide the fields into two contributions, ${\bf
B} = {\bf B}_1 + {\bf B}_2,\,  A = A_1 + A_2, \,
{\bf j} = {\bf j}_1 + {\bf j}_2,\, J = J_1 + J_2,$ etc., where
the subscript $n=1$ indicates that it is driven by the fluxoid
[driving term $D_1(r) = -\Phi_f /2\pi r$ and $A_1(r) = A_{J1}(r)$],
and the subscript $n =2$ indicates that it is driven by the
applied field [driving term
$D_2(r) = A_a(r)$ and $A_2(r) = A_a(r) + A_{J2}(r)$].
The London equation for contribution $n$ is
  \begin{equation} 
  J_n(r) = -[D_n(r) + A_{Jn}(r)]/\mu_0 \Lambda.
  \end{equation}

In this paper we will calculate the two contributions ($n$ = 1
and 2) to the magnetic flux $\Phi_n(a)$ through the hole,
total flux $\Phi_n(b)$ through the ring,  magnetic moment $m_n$,
and  total current $I_n$ around the ring using the definitions
  \begin{eqnarray}  
  \Phi_n(r) = 2\pi\! \int_0^r \!\! dr'\,r' B_n(r')=2\pi r
    A_n(r)\,,\\
  m_n = \pi\! \int_a^b \!\! dr\, r^2 J_n(r) , ~~~
  I_n = \! \int_a^b \!\! dr\, J_n(r) \,.
  \end{eqnarray}

The relation between the sheet current $J_n(r)$ and the vector
potential $A_{Jn}(r)$ it generates is obtained as follows. From the
Maxwell equation  $ \mu_0 {\bf j} = \nabla \times {\bf B} =
\nabla\times\nabla\times {\bf A} = -\nabla^2 {\bf A} $ (since here
$\nabla\cdot {\bf A} =0$) we obtain the 3D Biot-Savart law for the
current-generated part ${\bf A}_J({\bf r})$ of ${\bf A}$:
  \begin{equation}  
  {\bf A}_J({\bf r}) = \mu_0 \int d^3 r' {{\bf j(r')} \over 4\pi
  |{\bf r-r'}| }
  \end{equation} with ${\bf r} = (x,y,z)$. Integrating this over $z'$
and
$\varphi$, noting that ${\bf j}$ flows only inside the film
$-d/2 \le z \le d/2$, one obtains in the plane $z = 0$ for each component:
  \begin{equation}  
  A_{Jn}(r) = \mu_0 \int_a^b \! dr' J_n(r')\, Q(r,r')
  \end{equation} with the integral kernel
  \begin{equation}  
  Q(r,r') = \int_0^\pi \! {d\varphi \over 2\pi} {r'\cos\varphi
  \over (r^2 +r'^2 -2rr' \cos\varphi ) ^{1/2} } \,.
  \end{equation} This kernel may be written in terms of elliptic
integrals, but for transparency we prefer a fast direct numerical
integration of (7); see also Refs.~\onlinecite{2,3,4,5}. High
accuracy is achieved by substituting in the integral $\varphi
=\varphi(u) = \pi u -\sin(\pi u)$ and integrating over $0 \le u \le
1$ using an equidistant grid for $u$, $u_i = (i-1/2)/N_\varphi$,
$i=1,2,\dots, N_\varphi$, $N_\varphi \approx 30 - 60$, with weights
$w_i =\varphi'(u) /N_\varphi =[1-\cos(\pi u_i)] \pi/N_\varphi$:
  \begin{eqnarray}  
  Q(r,r') = \int_0^\pi \!\! d \varphi\, f(\varphi) =
  \int_0^1 \!\! du \,f[ \varphi(u) ]\, \varphi'(u)  \nonumber \\
  \approx  \sum_{i=1}^{N_\varphi}  f[\varphi(u_i)]\, w_i \,.
  \end{eqnarray}
Writing $J_n(r) = \int dr' J_n(r')
\delta(r-r')$ and inserting Eq.\ (2) into Eq.\ (6), we obtain
  \begin{equation}  
  D_n(r) = -\mu_0 \int_a^b\!\! dr' \,J_n(r')\,
   [\, Q(r,r') +\Lambda \delta(r-r')]\,.
  \end{equation}
(For introduction of finite $\lambda$ into other
geometries see Ref.~\onlinecite{14}.) Formally, the integral
equation (9) may be solved for the sheet current $J(r)$ by writing
  \begin{equation}  
  J_n(r) = -\mu_0^{-1} \!\int_a^b\!\! dr' \,D_n(r')\,K(r,r') \,,
  \end{equation}
where  $K(r,r')$
is the inverse of the kernel $ Q(r,r') +\Lambda \delta(r-r') $,
defined by
  \begin{equation}  
  \int_a^b\!\!\! dr' K(r,r') [Q(r'\!,r'')
    + \Lambda \delta(r'\! -\!r'')] = \delta(r\! -\!r'') \,.
  \end{equation}
The inverse kernel $K(r,r')$ is easily calculated
numerically by introducing an appropriate grid $r_i$ with weights
$w_i$ such that the integral is approximated by a sum,
  \begin{equation}  
  \int_a^b \! dr\, f(r) \approx \sum_{i=1}^{N_r} f(r_i)\, w_i \,.
  \end{equation} High accuracy is achieved in the present case,
where the integrated function $f(r)$ may have infinities at $r=a$ and
$r=b$, by a grid that is very dense near $r=a$ and $r=b$. A good
such grid is found by the substitution
$r=r(u)=a+(b-a)(10u^2 -15u^3 +6u^5)$, $r'(u)=30(b-a)(u-u^2)^2$,
yielding $r_i = r(u_i)$, $w_i = r'(u_i) / N_r$, with
$u_i = (i-1/2)/N_r$, $i=1,2, \dots, N_r$, $N_r \approx 30-100$. This
grid defines the vectors $J_{ni} =J_n(r_i)$, $D_{ni} =D_n(r_i)$,
 and the matrix $Q_{ij} = Q(r_i, r_j)$. Equation (9) then becomes a
sum (or matrix multiplication):
  \begin{equation}  
  D_{ni} = -\sum_{j=1}^{N_r}
   ( w_i Q_{ij}+\Lambda \delta_{ij} )\, \mu_0 J_{nj} \,.
  \end{equation}
This is inverted by
  \begin{equation}  
  \mu_0 J_{ni} = -\sum_{j=1}^{N_r} K_{ij} D_{nj} \,,
  \end{equation}
where $K_{ij}$ is an inverse matrix:
  \begin{equation}  
  K_{ij} = ( w_i Q_{ij} + \Lambda \delta_{ij} )^{-1}
  \end{equation}
(no summation over $i$; $\delta_{ij}=1$ if $i=j$;
otherwise
$\delta_{ij}=0$).
The matrix equation from which the total sheet current
is determined for given $B_a$ and $\Phi_f$  thus reads explicitly:
  \begin{equation}  
  \mu_0 J(r_i) = -\sum_{j=1}^{N_r} K_{ij}\, \Big( \frac{r_j}{2}
   B_a - \frac{\Phi_f}{ 2\pi r_j} \Big) \,.
  \end{equation} Finally, one hint is required without which this
method may not work or is inaccurate. The matrix $Q_{ij}$ has
infinite diagonal terms $Q_{ii}$, since $Q(r,r')$, Eq.~(7), diverges
logarithmically when $r$ approaches $r'$. Namely, one has
\begin{eqnarray}  
  Q_{i,i+1} \approx Q(r_i, r_i+w_i) \approx
  {1 \over 2\pi} \ln {r_i \over w_i} \,.
  \end{eqnarray} This problem was dealt with in detail in Refs.\
\onlinecite{2} and \onlinecite{3}, where the optimum choice of the
$Q_{ii}$ was found first numerically \cite{2} and then analytically
\cite{3}, e.g., from the condition that an infinite disk ideally
screens two coils separated by the disk. One thus has the complete
definition of the matrix $Q_{ij}$:
  \begin{eqnarray}  
  Q_{ij} &=& Q(r_i, r_j),~~ i \ne j \,, \nonumber \\
  Q_{ii} &=& {1 \over 2\pi}
    \Big( \ln {16\pi r_i \over w_i} -2 \Big) \,.
  \end{eqnarray} In Eq.~(15) the diagonal term is $w_i Q_{ii} +
\Lambda$. Thus, when $\Lambda$ is larger than the maximum value of
the weight $w_i$ (or the spacing between grid points), which is of
order $(b-a)/N_r$, then the choice of the $Q_{ii}$ is not critical,
and for the computation of $J(r)$ one may even put $Q_{ii}=0$. For
small $\Lambda$, however, the correct choice of $Q_{ii}$ is
important.

 \begin{figure}  
\includegraphics[scale=.48]{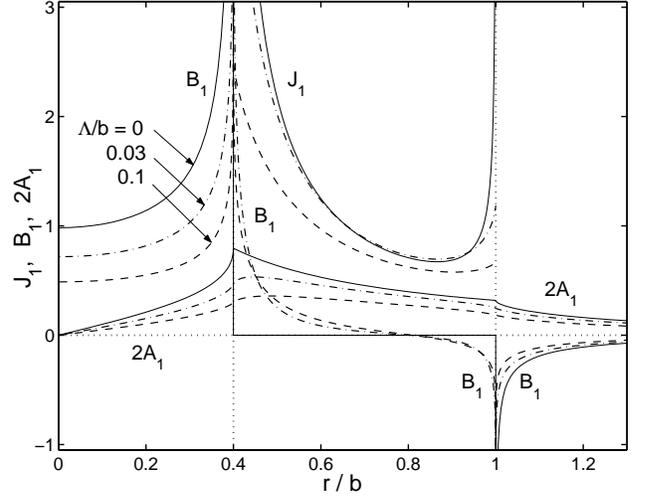}
\caption{\label{fig1} Profiles of sheet current $J_1(r)$ in the
ring,  perpendicular magnetic induction
$B_1(r)=(1/r)(rA_1)'$, and  current-generated vector potential
$A_1(r)= A_{J1}(r)$ in the plane of the ring for the case of trapped
flux and zero applied field ($B_a=0$, $\Phi_f >0$, Figs.~1--6). The
hole radius is $a =0.4 b$ and the 2D penetration depth is
$\Lambda = 0$ (solid curves), $\Lambda=0.03b$ (dot-dashed), and
$\Lambda= 0.1b$ (dashed). Plotted are the dimensionless quantities
$J_1/(\Phi_f/\mu_0 b^2)$, $B_1/(\Phi_f/b^2)$, and $2A_1/(\Phi_f/b)$.
 }
 \end{figure}     

 \begin{figure}   
\includegraphics[scale=.48]{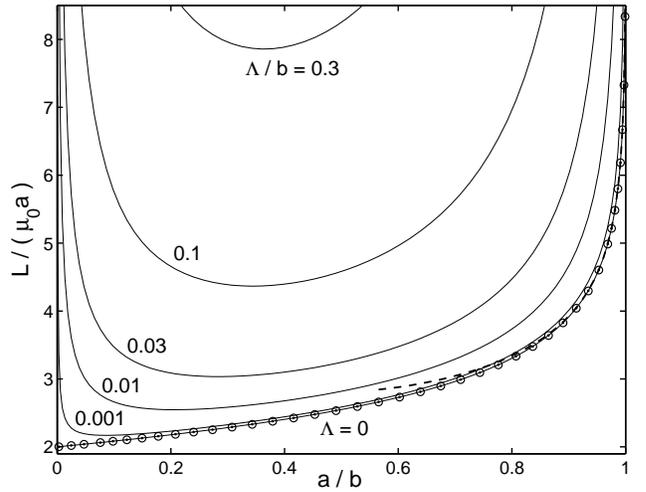}
\caption{\label{fig2} Computed self-inductance $L =\Phi_f/I_1$ of a
superconducting thin flat ring with hole radius $a$ and outer radius
$b$, plotted in units of  $\mu_0 a$ for 2D penetration depths
$\Lambda = 0.001$, 0.01, 0.03, 0.1, and 0.3 in units of $b$. Note
that even very small $\Lambda \ll b$ considerably enhances the
inductance of rings with a small central hole ($a/b \ll 1$) as
compared with the case $\Lambda=0$. The circles depict the empirical
expression (20) valid for
$\Lambda=0$; they perfectly coincide with the dots that mark the
curve computed for $\Lambda=0$. The dashed curve shows the limiting
expression $L_1$, Eq.~(19). See also Fig.~6 below, showing the
inverse self-inductance $\mu_0 a /L$.
 }
 \end{figure}    

\section{Self-inductance of flat rings}  

   According to Eq.~(16) the current in a ring originates
 from either an applied field $B_a$,  trapped flux related to the
parameter $\Phi_f$, or  both. In this section we put $B_a=0$ and
compute the sheet current $J_1(r)$ for finite trapped fluxoid
$\Phi_f$. The total current $I_1$ is then given by Eq.~(4). The
current generates a flux $\Phi_1(r)=2\pi r A_1(r)$, Eq.~(3),
with $A_1(r) = A_{J1}(r)$ from Eq.~(6).
  Figure 1 shows the profiles of the sheet current $J_1(r)$,
perpendicular induction $B_1(r) = (1/r)(rA_1)'$, and vector
potential $A_1(r)$ for a ring with $a/b= 0.4$ for three
$\Lambda/b$ values (0, 0.03, 0.1) for the case of flux trapping
with $B_a=0$ and $\Phi_f >0$. One can see that finite $\Lambda$
removes the infinities of $J_1(r)$ at $r=a$ and $r=b$
[$J_1 \propto (r-a)^{-1/2}$ and $J_1 \propto (b-r)^{-1/2}$;
see Ref.~\onlinecite{1}], which lead to
a similar infinity of $B_1(r)$. For $\Lambda >0$, $J_1(a)$ and
$J_1(b)$ are finite, and $B_1(r)$ penetrates the superconductor but
still exhibits a logarithmic infinity at $r=a$ and $r=b$, which is
caused by the abrupt jump of $J_1(r)$ to zero. Note also that with
increasing $\Lambda$ the total current $I_1$, the magnetic flux
$\Phi_1(a)$ in the hole, and the field $B_1$ in the hole {\it
decrease}, while the London fluxoid
$2\pi r [A_1(r)+\mu_0 \Lambda J_1(r)] =\Phi_f$ remains
constant.\cite{15}

  In the case of ideal screening, i.e. for $\Lambda=0$, no magnetic
field  penetrates the ring material, and thus the flux
$\Phi_1(r)$ is the same for any $r$ between $r=a$ and $r=b$. The
self-inductance of the ring may then be defined as
$L=\Phi_1/I_1$ with $\Phi_1 = \Phi_1(a) = \Phi_1(b)$. For small
reduced inner radius $\tilde a \equiv a/b \ll 1$ the flux trapped
in the hole was found by Clem\cite{1,11} to be $\Phi_1(a) = 2\mu_0
aI_1$; thus the inductance
$L=\Phi_1/I_1$ approaches $L_0 = 2\mu_0 a$. In the opposite limit,
for narrow rings with $\tilde a \to 1$, $L$ approaches\cite{1,5}
  \begin{eqnarray}  
  L_1 &=& \mu_0 R \,[\, \ln(8R/w) -2+\ln4 \,] \nonumber \\
  &=& \mu_0 b (1+\tilde a)(\,\tanh^{-1} \tilde{a} -1 +\ln4\,)
  \end{eqnarray} [$R=(a+b)/2$, $w=b-a$]. For arbitrary $\tilde a$,
but still
$\Lambda=0$, $L$ was computed in Ref.~\onlinecite{1}, where a useful
empirical formula was presented,
  \begin{eqnarray}  
  L_2=\mu_0 b \, [\,\tilde{a} -\!0.197 \tilde{a}^2 \!-\!0.031
  \tilde{a}^6 \!+\!(1\!+\!\tilde{a}) \tanh^{-1}\!\tilde{a}\,]\,.
  \end{eqnarray} This fit is confirmed by our method, as shown in
Fig.~2; its relative deviation from the exact $L$ ranges from
$-0.005\%$  to $+0.06\%$.

 \begin{figure}   
\includegraphics[scale=.48]{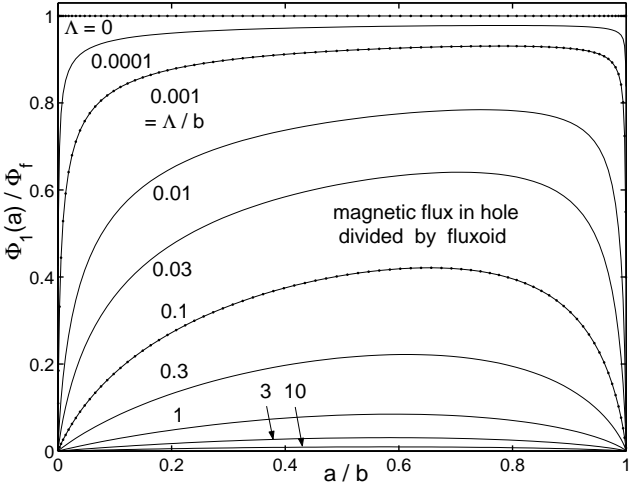}
\caption{\label{fig3} Magnetic flux $\Phi_1(a)$ in the hole of a
superconducting thin flat ring with hole radius $a$ and outer radius
$b$ for the trapped-flux case as in Figs.~1--6 (applied field
$B_a=0$ and fluxoid $\Phi_f >0$) for 2D penetration depths
$\Lambda/b =0$, 0.0001, 0.001, 0.01, 0.03, 0.1, 0.3, 1, 3, and 10.
  The dots on some of the curves mark the $a/b$ grid used here
  in all such figures.
 }
 \end{figure}    

 \begin{figure}   
\includegraphics[scale=.48]{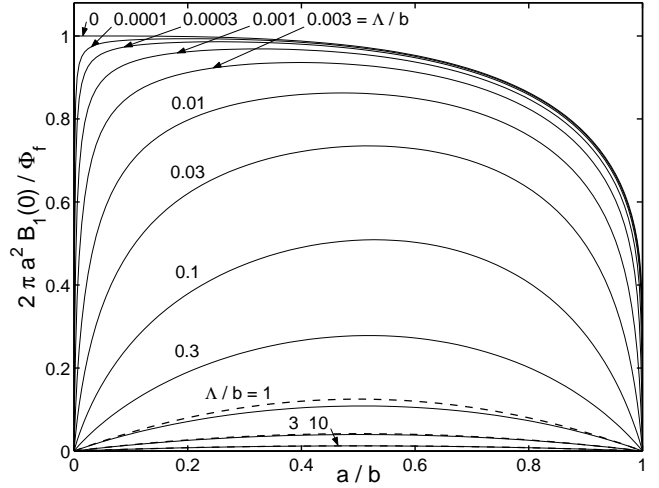}
\caption{\label{fig4} Minimum of the magnetic field $B_1(0)$
occurring in the center of the ring for the trapped-flux case
($B_a=0$ and $\Phi_f >0$) plotted as $2\pi a^2 B_1(0) / \Phi_f$ for
$\Lambda/b = 0$, 0.0001, 0.0003, 0.001, 0.003, 0.01, 0.03, 0.1, 0.3,
1, 3, and 10. The dashed curves show the large-$\Lambda$
approximation, Eq.~(41), for $\Lambda/b = 1$, 3, 10.
 }
 \end{figure}    

 \begin{figure}   
\includegraphics[scale=.48]{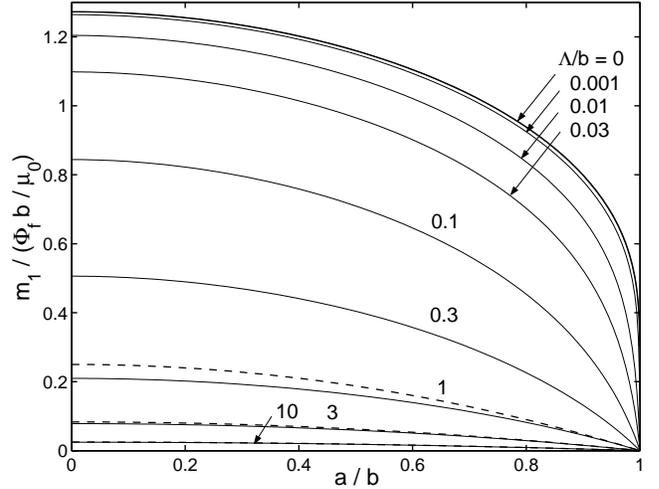}
\caption{\label{fig5} Magnetic moment $m_1$ of a superconducting
thin flat ring with inner and outer radii $a$ and $b$ for the
trapped-flux case ($B_a=0$ and $\Phi_f >0$) plotted as the
dimensionless ratio $\alpha_m =\mu_0 m_1 /b\Phi_f$ for
$\Lambda/b =0$, 0.001, 0.01, 0.03, 0.1, 0.3, 1, 3, and 10.
The dashed curves show the large-$\Lambda$ approximation, Eq.~(42),
for $\Lambda/b = 1$, 3, 10. See also Fig.~12 below.
 }
 \end{figure}    

 \begin{figure}   
\includegraphics[scale=.48]{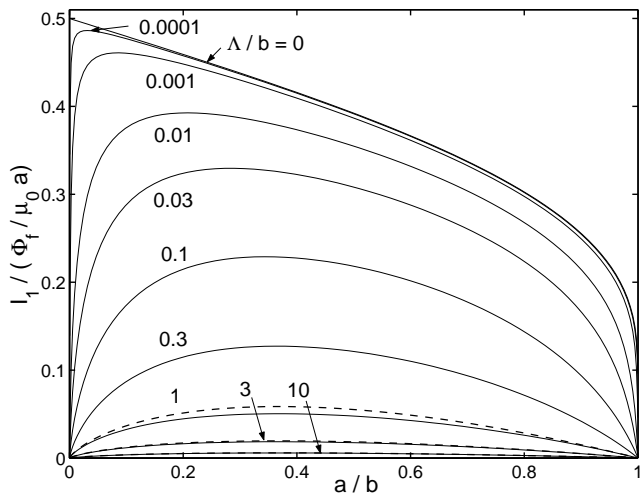}
\caption{\label{fig6} Total current  $I_1$ in the flat ring with
trapped flux ($B_a=0$ and $\Phi_f >0$) plotted as the dimensionless
ratio $\alpha_I = \mu_0 a I_1 /
\Phi_f=\mu_0 a/ L$ for $\Lambda/b=0$, 0.0001, 0.001, 0.01, 0.03, 0.1,
0.3, 1, 3, and 10. Same data as in Fig.~2 but inverted.
The dashed curves show the large-$\Lambda$ approximation, Eq.~(40),
for $\Lambda/b = 1$, 3, 10.
 }
 \end{figure}    

 When the effective penetration depth $\Lambda$ is finite, the
magnetic field penetrates into the ring material, and the magnetic
flux $\Phi_1(r)$ is no longer constant when $r$ changes from $r=a$
to
$r=b$. The definition of $L =\Phi_1/I_1$ via magnetic flux is thus
not unique; in particular,
$\Phi_1(a)/I_1 \ne \Phi_1(b)/I_1$ when $\Lambda > 0$. In this
general case one must use the definition of $L$ via the
electromagnetic energy of the ring, $ E= (1/2) L I_1^2$. This energy
is composed of the magnetic energy $E_m$ and the kinetic energy
$E_k$ of the currents, as is evident from the energy integral of
London theory:
  \begin{eqnarray}  
  E=E_m\!+\!E_k={\mu_0 \over 2}\int\! d^3r (\, {\bf H}_1^2
                          +\lambda^2 {\bf j}_1^2 \,)
  \end{eqnarray} with ${\bf H}_1 = \mu_0^{-1} {\bf B}_1$ and
${\bf j}_1= \nabla \times {\bf H}_1$. The integral (21) over all
space can be transformed into an integral over the superconductor by
introducing the vector potential. For a flat ring this yields
 \begin{eqnarray}  
 E=E_m \!+\!E_k=\pi\int_a^b \!\!dr\, r [\,J_1(r)A_1(r)
 +\mu_0\Lambda J_1(r)^2\,]\,.
 \end{eqnarray}
From the two energy terms when $B_a = 0$ one may
define the geometric inductance $L_m = 2E_m/I_1^2$ and the kinetic
inductance
$L_k = 2E_k/I_1^2$, yielding the total self-inductance
$L = L_m +L_k = 2E /I_1^2$. The energy $E$, Eq.~(22), may also be
written as
 \begin{eqnarray}  
 E= E_m +E_k = (1/2)\, \Phi_f I_1 \,,
 \end{eqnarray} where $\Phi_f = 2\pi r (A_1 +\mu_0 \Lambda J_1) =$
const (for
$a \le r \le b$) is the London fluxoid.\cite{13,15} We may derive
Eq.\ (23) from (22) by noting that the constant combination  $2\pi
r (A_1 +\mu_0 \Lambda J_1) =\Phi_f$ can be factored out of the
integrand.

  From Eq.~(23) and the definition $E=(1/2)L I_1^2$ it immediately
follows that
\begin{equation} 
L = \Phi_f / I_1.
\end{equation}
This general result, valid for arbitrary penetration
depth $\Lambda$, differs from the definition used previously\cite{1}
for $\Lambda=0$, in that the magnetic flux
$\Phi_1(a)$ through the hole is replaced by the fluxoid $\Phi_f$,
which  coincides with the flux
$\Phi_1(a)$ in the hole only in the special case $\Lambda=0$.

  Equation (23) also may be derived by considering how the fluxoid
in the ring may be increased from zero to $\Phi_f$ by moving
vortices (Pearl vortices \cite{16,17,18} of short length $d$) from
the outer radius through the ring into the hole. Each vortex has to
cross the current-carrying ring, where a Lorentz force $\phi_0
J_1(r)$ acts on it. Integrating this force from
$r=b$ to $r=a$, one obtains the energy $\phi_0 I_1$. Each crossing
vortex increases the phase change of the superconductor order
parameter around a circle in the ring by $2\pi$ and thus increases
the fluxoid by $\phi_0$. Noting that the total current $I_1$ is
proportional to the number of vortices that already are in the hole,
one obtains
$E = (1/2) \Phi_f I_1$, Eq.~(23).

  Figure 2 shows the inductance $L = \Phi_f/I_1$ for various ratios
$\Lambda/b = 0$, 0.001, 0.01, 0.03, 0.1, and 0.3. The circles show
the fit (20), which is an excellent approximation for
$\Lambda=0$ and all hole radii $a$. Note that $L$ increases with
increasing penetration depth $\Lambda$. Even small
$\Lambda/b=0.001$ noticeably enhances $L$ of rings with a small
hole.
Figure 3 shows the magnetic flux in the hole, $\Phi_1(a)=2\pi
aA_1(a)$, referred to the fluxoid  $\Phi_f$ trapped in the ring.
Note that
$\Phi_1(a)$ may be much smaller than $\Phi_f$ even for small
$\Lambda/b$. A similar plot, Fig.~4, shows the minimum field in the
hole, $B_1(0)$, occurring at the center $r=0$ (cf.\ Fig.~1) plotted
as
$2\pi a^2 B_1(0)/\Phi_f$. These curves look qualitatively similar to
$\Phi_1(a)/\Phi_f$ in Fig.~3. For
$\Lambda=0$ and $a\ll b$ one has $2\pi a^2 B_1(0)/\Phi_f \to 1$.

 The magnetic moment $m_1$ of this ring (still for $B_a=0$ and
$\Phi_f >0$) is depicted in Fig.~5 as the dimensionless ratio
$\alpha_m = \mu_0 m_1 /b\Phi_f$; for further approximations to these
curves see the dotted and dot-dashed curves in the similar Fig.~12 below.
The total current $I_1$ times the inner radius $a$ is depicted in
Fig.~6 as the dimensionless ratio $\alpha_I = \mu_0 a I_1/
\Phi_f$. Actually Fig.~6 shows the same data as Fig.~2, but
inverted, since the plotted quantity $\mu_0 a I_1 /\Phi_f$ equals
$\mu_0 a/ L$; however, Fig.~6 shows the entire range
$0\le \Lambda \le \infty$.

 \begin{figure}  
\includegraphics[scale=.48]{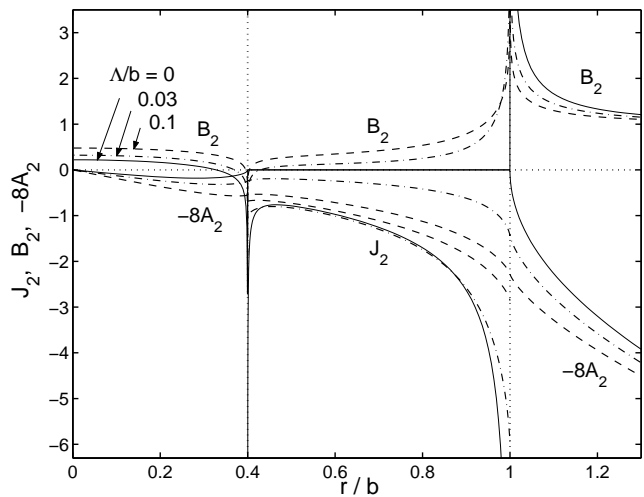}
\caption{\label{fig7} Profiles of sheet current $J_2(r)$ in the
ring, perpendicular magnetic induction $B_2(r)=(1/r)(rA_2)'$, and
vector potential $A_2(r)$ in the plane of the ring for the
zero-fluxoid state in a finite applied field
 ($B_a>0$, $\Phi_f=0$, Figs.~7-12). The hole radius is $a =0.4 b$
and the 2D penetration depth is
$\Lambda = 0$ (solid curves), $\Lambda=0.03 b$ (dot-dashed), or
$\Lambda= 0.1 b$ (dashed), as in Fig.~1. Plotted are the
dimensionless quantities  $J_2/H_a$, $B_2/B_a$, and  $-8A_2/bB_a$.
 }
 \end{figure}     

 \begin{figure}   
\includegraphics[scale=.48]{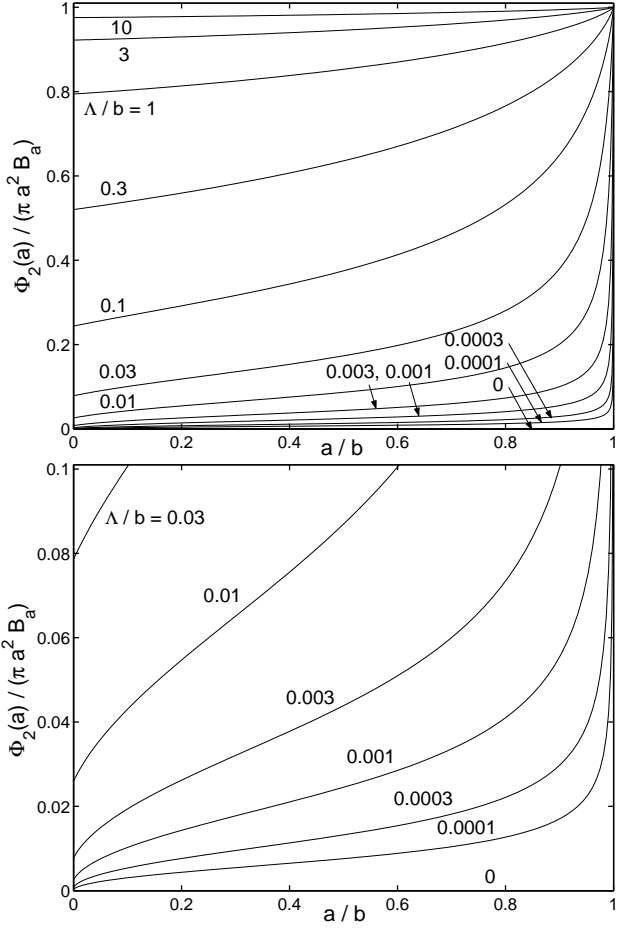}
\caption{\label{fig8} Magnetic flux $\Phi_2(a)$ in the hole of a
flat ring when $B_a>0$ and $\Phi_f=0$ in units of $\pi a^2 B_a$ for
$\Lambda/b =0$, 0.0001, 0.0003, 0.001, 0.003, 0.01, 0.03, 0.1, 0.3,
1, 3, and 10. The bottom plot shows the same data ten times enlarged.
 }
 \end{figure}    

 \begin{figure}[t]  
\includegraphics[scale=.48]{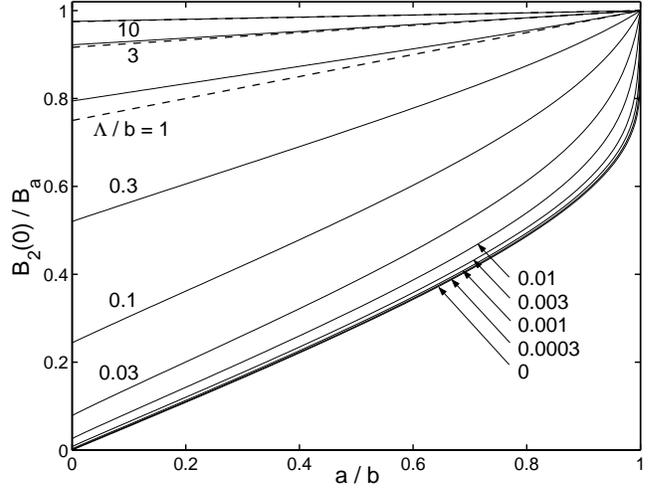}
\caption{\label{fig9} Magnetic field $B_2(0)$ at the center
$r=0$ of a thin flat ring when $B_a >0$ and $\Phi_f=0$ in units
$B_a$ for $\Lambda/b =0$, 0.0003, 0.001, 0.003, 0.01, 0.03, 0.1,
0.3, 1, 3, and 10. The dashed curves show the large-$\Lambda$
approximation, Eq.~(43), for $\Lambda/b = 1$, 3, 10.
 }
 \end{figure}    

 \begin{figure}[t]  
\includegraphics[scale=.48]{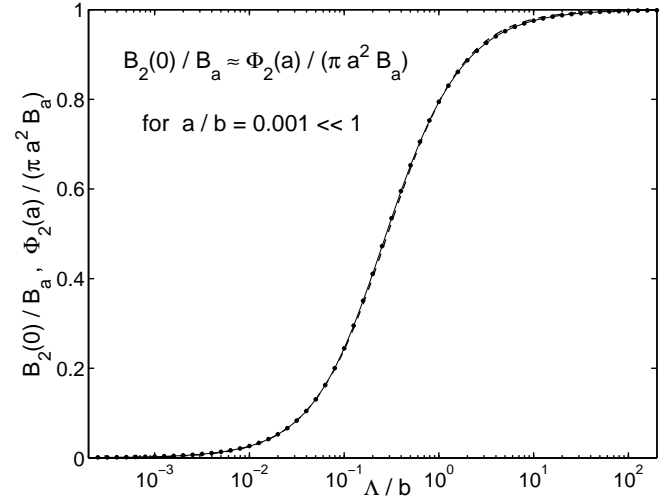}
\caption{\label{fig10} Penetrated flux in thin flat rings with small
hole radius $a\ll b$ for $B_a >0$, $\Phi_f =0$, plotted versus
$\Lambda/b$ as $\Phi_2(a)/\pi a^2 B_a$ (dots), which in this limit
is nearly equal to $B_2(0)/B_a$ (solid curve) and is well
approximated by Eq.~(25), shown as the dashed curve.
 }
 \end{figure}    

 \begin{figure}   
\includegraphics[scale=.48]{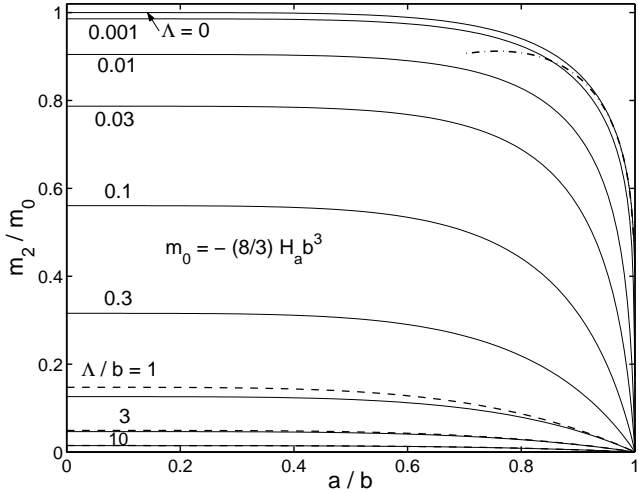}
\vspace{-1mm}
\caption{\label{fig11} Magnetic moment $m_2$ of a thin flat ring
when $B_a >0$ and $\Phi_f=0$, plotted as the dimensionless ratio
$\beta_m = m_2/m_0$, where $m_0 = -(8/3)H_a b^3$ for
$\Lambda/b =0$, 0.001,  0.01, 0.03, 0.1, 0.3, 1, 3, and 10. The
dot-dashed curve shows the limit $b \to a$, Eq.~(26).
The dashed curves show the large-$\Lambda$ approximation, Eq.~(44),
for $\Lambda/b = 1$, 3, 10.
 }
 \end{figure}    

 \begin{figure}   
\includegraphics[scale=.48]{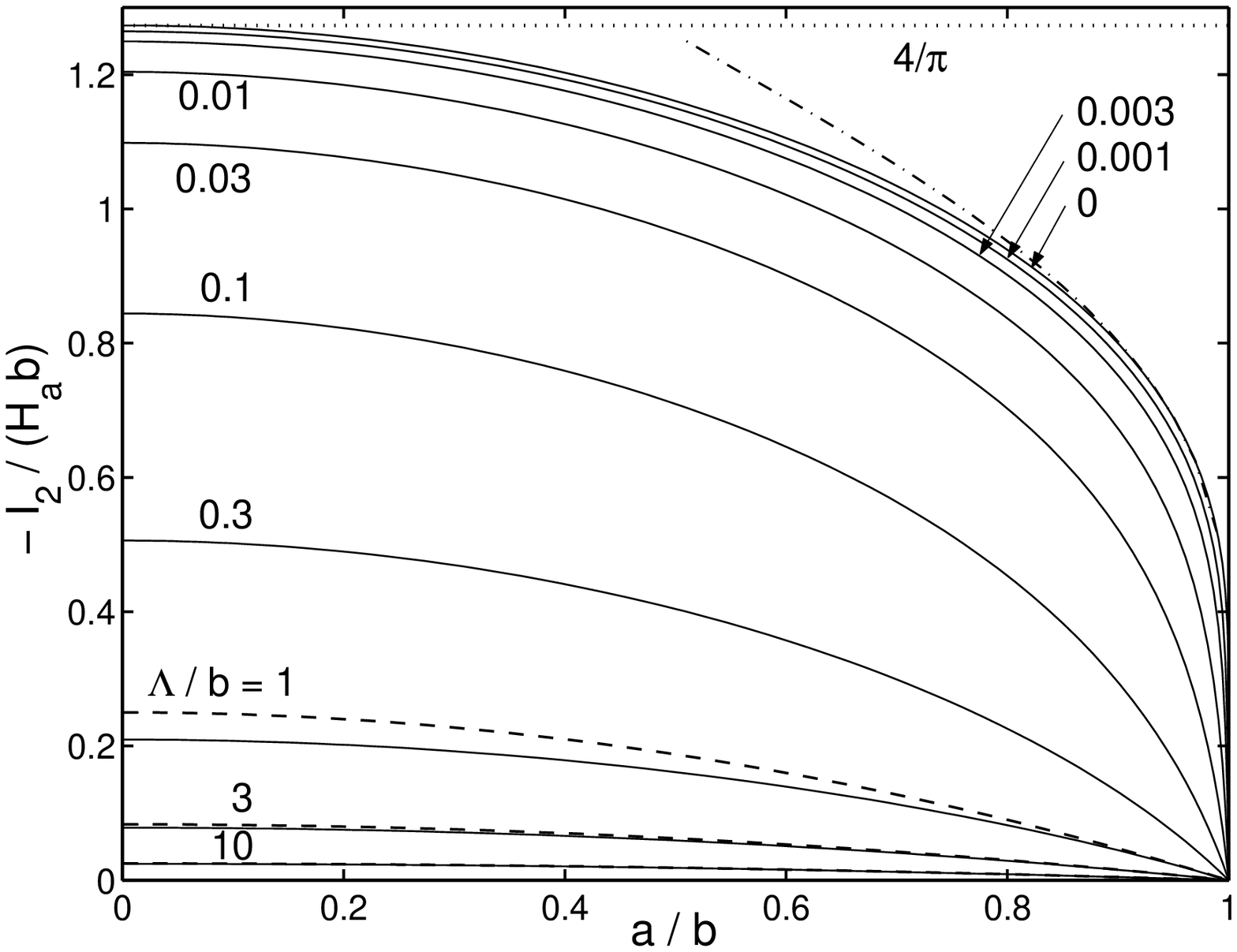}
\vspace{-1mm}
\caption{\label{fig12} Total current $I_2$ in a thin flat ring when
$B_a >0$ and $\Phi_f=0$ for
$\Lambda/b =0$, 0.001,  0.003, 0.01, 0.03, 0.1, 0.3, 1, 3, and 10,
plotted as the dimensionless ratio $\beta_I = -I_2/bH_a$.
The dotted and dot-dashed curves show the limits, Eq.~(27).
Note that the curves of $\beta_I = -I_2/bH_a$  agree exactly with
those in Fig.~5, $\alpha_m = \mu_0 m_1/b\Phi_f$, although they
describe different physical quantities. The dashed curves show the
large-$\Lambda$ approximation, Eq.~(45), for $\Lambda/b = 1$, 3, 10.
 }
 \end{figure}    

\section{Zero-fluxoid state}  

  This section considers the zero-fluxoid state reached when  a
thin superconducting  ring is slowly cooled in zero field. A
magnetic field $B_a=\mu_0 H_a$ is then applied, but the fluxoid
$\Phi_f$ in the ring then remains zero. When $\Lambda=0$, one has
ideal screening, such that the magnetic flux $\Phi(a)$ in the hole
is zero. For finite $\Lambda >0$ the screening is incomplete, some
flux leaks through the ring, and the flux
$\Phi(a)$ in the hole is no longer zero. In this section we
therefore put $B_a >0$ and $\Phi_f=0$ in our numerics. Figure 7
shows for this case the profiles of $J_2(r)$, $B_2(r)$, and $A_2(r)$
for a ring with $a/b=0.4$ for $\Lambda/b = 0$, 0.03, and 0.1. Note
that the induction $B_2(r)$ changes sign inside the hole and has a
negative infinity at the inner edge
$r=a$ and a positive infinity at the outer edge $r=b$. For
$\Lambda=0$ the integral of $B_2(r)$ over the hole area is
$\Phi_2(a)=0$, but for $\Lambda>0$ the flux $\Phi_2(a) >0$.

  Figure 8 shows the penetrated flux $\Phi_2(a)$ in units of its
maximum value $\pi a^2 B_a$ reached in the limit of
$\Lambda \gg b$, and Fig.~9 shows the field maximum $B_2(0)$
occurring at the center $r=0$ of the ring. Both $\Phi_2(a)$ and
$B_2(0)$ increase monotonically with both $a$ and $\Lambda$, but
while $B_2(0)$ at small
$a$ increases linearly with the radius $a$, the penetrated flux
$\Phi_2(a)$ at small $a$ and small $\Lambda$ has negative curvature;
see bottom of Fig.~8. The penetrated flux $\Phi_2(a)/\pi a^2 B_a
\approx B_2(0)/B_a$ in the limit of small hole radius $a/b \ll 1$ is
depicted in Fig.~10 as a function of $\Lambda/b$. This curve is well
fitted by
  \begin{eqnarray}  
  {\Phi_2(a) \over \pi a^2 B_a} \approx {B_2(0) \over B_a}
  \approx {1\over 2}\!+\!{1\over 2} \tanh\! \Big( 2.88
  \ln{\Lambda\over b} - \! 0.675 \Big) \,.
  \end{eqnarray}  

Figure 11 shows the magnetic moment $m_2$ of the ring in units of
$m_0 = -(8/3)b^3 H_a$, which is reached for ideally
screening disks ($a=\Lambda=0$);\cite{2,Clem94}
the dimensionless ratio plotted is $\beta_m = m /m_0.$  For
$\Lambda=0$ and $a \to b$ one has the limit\cite{5}
  \begin{equation}  
  \frac{m_2}{m_0}  = \frac{3\pi^2}{128}\frac{( 1+a/b)^3}{
         \tanh^{-1} (a/b) -1+\ln 4} .
  \end{equation} Figure 12 shows the total current  $I_2$ in
the ring induced by the applied field, expressed in terms of the
dimensionless ratio $\beta_I = -I_2/bH_a$. The limits for
$\Lambda=0$ are\cite{1,5}
  \begin{eqnarray}  
  I_2 =& -(4 /\pi) \,b\,H_a~ & {\rm for~} a\to 0 \,,\nonumber \\
 I_2 =& {\displaystyle   
      - {(\pi/4)\, (a+b) \, H_a \over \tanh^{-1} (a/b)
        -1\! +\ln 4} }~ & {\rm for~} a\to b \,.
  \end{eqnarray} Note that the curves in Fig.~12 exactly coincide
with the curves in Fig.~5, even though they depict different
physical quantities for different cases. This identity,
$\alpha_m = \beta_I$, can be proved by evaluating the sum of energy
integrals $F_{12} = \int\! d^3r \, {\bf B}_1 \cdot {\bf B}_2/\mu_0
+\int\! d^3r \mu_0 \lambda^2 {\bf j}_1 \cdot {\bf j}_2 \,$,
where the first integral extends over all space, including the coil
producing the applied field $B_a$, and the second integral extends
only over the volume of the ring.
With the help of the vector identity $\nabla \cdot ({\bf A} \times
{\bf B}) = {\bf B} \cdot \nabla \times {\bf A}
-{\bf A} \cdot (\nabla \times {\bf B})$ with ${\bf A} = {\bf
A}_2$ and  ${\bf B} = {\bf B}_1$,  the divergence theorem,
and  Eq.~(2) with $n = 2$, we can show that
$F_{12} = 0$.   Then, using the same vector identity but with
${\bf A} = {\bf A}_1$ and  ${\bf B} = {\bf B}_2$, the divergence
theorem,  Ampere's law, and Eqs.~(2), (4), (6), and (7) with
$n= 1$, we obtain $ F_{12}  = m_1B_a +\Phi_fI_2 = 0$, which yields
$\alpha_m = \mu_0m_1/b\Phi_f = -\mu_0I_2/bB_a = \beta_I$.  This also
can be proved by inserting $J_1$ and $J_2$ of Eq.~(10) into the
definitions (4) of $m_1$ and $I_2$, renaming the variables
$r \leftrightarrow r'$, and noting the symmetry
of $K(r,r') =K(r'\!,r) (r'\!/r)$ defined by Eq.~(11). This symmetry
follows from the symmetry of $Q(r,r')=Q(r'\!,r)(r'\!/r)$ defined by
Eq.~(7), from Eq.~(11) noting that
$(r''\!/r)\,\delta(r-r'') = \delta(r-r'')$, and from the additional
property of the inverse kernel that
$\int_a^b\! dr' [\,Q(r''\!,r')  + \Lambda
 \delta(r''\! -r')]K(r'\!,r) = \delta(r''\! - r)$.

 \begin{figure}   
\includegraphics[scale=.48]{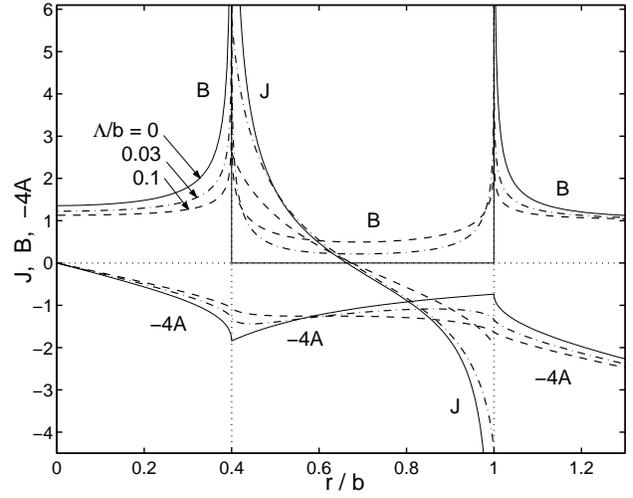}
\vspace{-1mm}
\caption{\label{fig13}  Profiles of sheet current $J(r)$ in the
ring, perpendicular magnetic induction $B(r)=(1/r)(rA)'$m and vector
potential $A(r)$ in the plane of the ring for the case of zero total
current $I=0$ (flux focusing, Figs.~13-17). The hole radius is $a
=0.4 b$ and the 2D penetration depths are
$\Lambda = 0$ (solid curves), $\Lambda=0.03 b$ (dot-dashed), or
$\Lambda= 0.1 b$ (dashed), as in Figs.~1 and 7. Plotted are the
dimensionless quantities  $J/H_a$, $B/B_a$, and  $-4A/b B_a$.
 }
 \end{figure}    

 \begin{figure}  
\includegraphics[scale=.48]{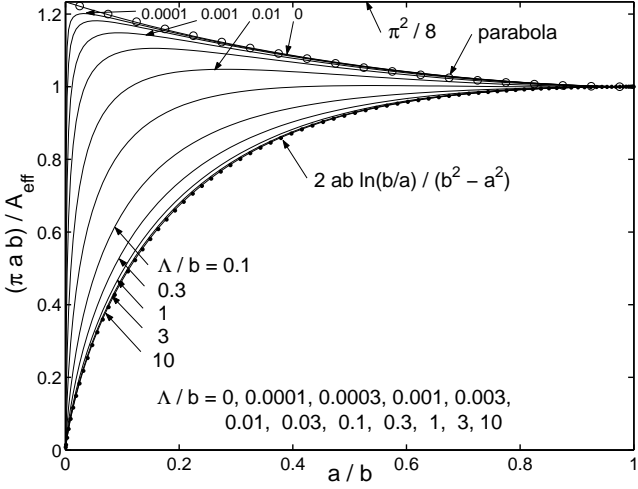}
\caption{\label{fig14} Reciprocal of the effective area
$A_{\rm eff} = \Phi_f / B_a$ of a thin flat ring with $I=0$ (flux
focusing) plotted as $(\pi a b) /A_{\rm eff}$ versus the radius
ratio $a/b$ for penetration depths $\Lambda/b = 0$, 0.0001, 0.0003,
0.001, 0.003, 0.01, 0.03, 0.1, 0.3, 1, 3, and 10. The circles mark
the parabolic fit, Eq.~(29), and the dots nearly coinciding with
the curves $\Lambda/b = 3$ and 10 show the limit
$\Lambda \to \infty$, Eqs.~(30), (47).
 }
 \end{figure}    

 \begin{figure}  
\includegraphics[scale=.48]{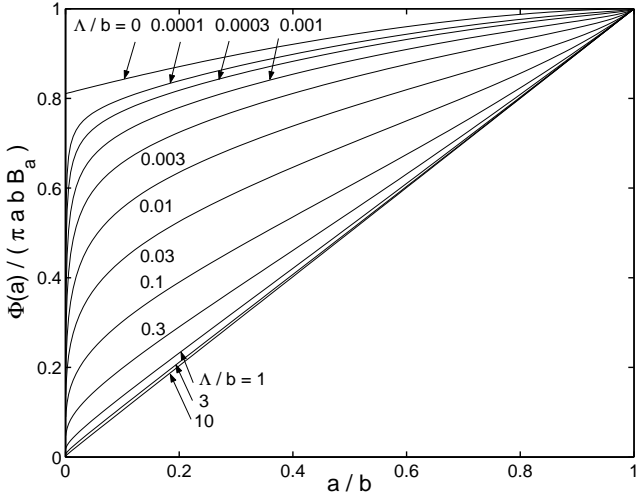}
\caption{\label{fig15}  Magnetic flux $\Phi(a)$ in the hole of a
thin flat ring with $I=0$ (flux focusing) plotted as
$\Phi(a)/\pi a b B_a$ versus $a/b$ for  $\Lambda/b = 0$, 0.0001,
0.0003, 0.001, 0.003, 0.01, 0.03, 0.1, 0.3, 1, 3, and 10.
 }
 \end{figure}    

 \begin{figure}  
\includegraphics[scale=.48]{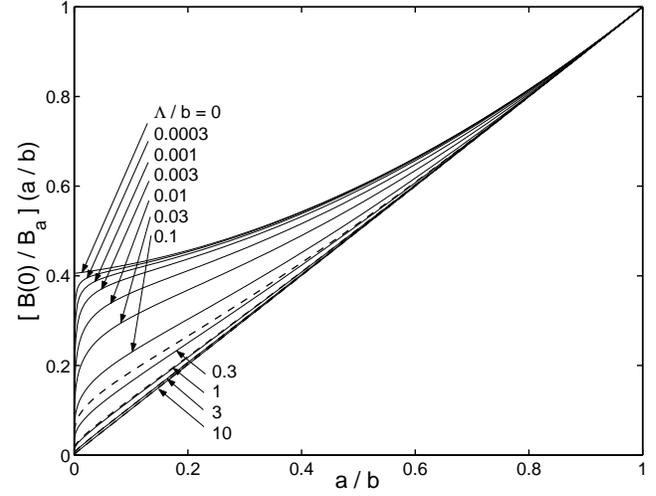}
\caption{\label{fig16} Magnetic induction $B(0)$ in the center of a
thin flat ring with $I=0$ (flux focusing) plotted as
$[B(0)/B_a] (a/b)$ versus $a/b$ for  $\Lambda/b = 0$,
 0.0003, 0.001, 0.003, 0.01, 0.03, 0.1, 0.3, 1, 3, and 10.
The dashed curves show the large-$\Lambda$ approximation, Eq.~(48),
for $\Lambda/b = 0.3$, 1, 3, 10.
 }
 \end{figure}    

 \begin{figure}  
\includegraphics[scale=.48]{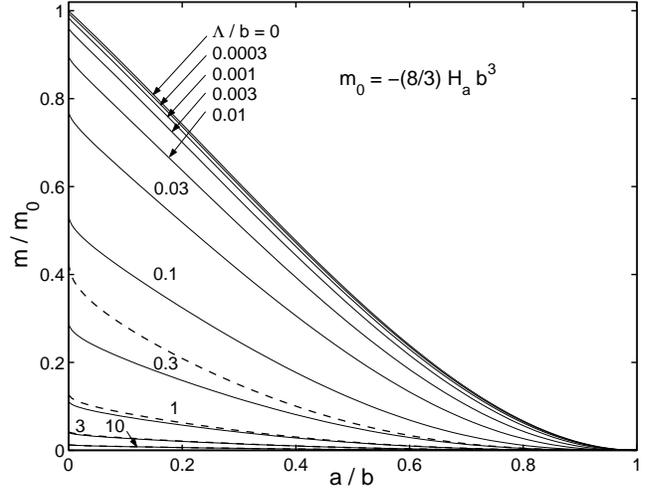}
\caption{\label{fig17} Magnetic moment $m$ of a thin flat ring when
$I=0$ (flux focusing) in units of $m_0 = -(8/3)H_a b^3$ for
$\Lambda/b =0$, 0.0003, 0.001, 0.003, 0.01, 0.03, 0.1, 0.3, 1, 3,
and 10. The dashed curves show the large-$\Lambda$ approximation,
Eq.~(49), for $\Lambda/b = 0.3$, 1, 3, 10.
 }
 \end{figure}    

\section{Flux Focusing}  

  This section considers a ring that is in a perpendicular magnetic
field $B_a$ and contains a fluxoid
$\Phi_f$ chosen such that no net current circulates around the ring.
This circularly symmetric situation approximates a ring with a
narrow slit along one of its radii as proposed by Clem.\cite{11,1}
It neglects both the magnetic flux in the slit and the effects of
the radial currents that flow in opposite directions alongside the
slit. The slit interrupts the current I and allows magnetic flux to
penetrate into the hole such that the condition
$I=0$ holds. When
$B_a$ is increased with ramp rate $dB_a / dt$, a voltage $U$ appears
across the slit at radius r,
$U = -d\Phi_f / dt = -(dB_a / dt) A_{\rm eff}$, where
$A_{\rm eff} = \Phi_f / B_a$ is the effective area of the slitted
ring. In our approximation of a circularly symmetric ring the
fluxoid $\Phi_f$ is given by
$\Phi_f =2\pi r [\,(r/2)B_a + A_J(r) + \mu_0 \Lambda J(r) \,]$ for
any $r$ in the superconductor, $a\le r \le b$, and it equals the
parameter $\Phi_f$ entering Eq.~(2). In a superconductor without a
slit, $\Phi_f = N \phi_0$ is quantized, with integer $N$. When the
ring has a radial slit, it may be used to construct a dc SQUID by
connecting the two banks of the slit to a superconducting current
lead via two identical small Josephson junctions, each with maximum
supercurrent $I_0$.\cite{12,Barone}  The critical current of the resulting
dc SQUID is
  \begin{equation}  
  I_c = 2 I_0 |\cos(\pi B_a A_{\rm eff} / \phi_0) | \,,
  \end{equation} where the effective area is
$A_{\rm eff} = \Phi_f / B_a$. Note that the effective area can be
calculated as $A_{\rm eff} =
\Phi(a) / B_a$ (as in Ref.\ \onlinecite{1}) only in the limit
$\Lambda=0$, when the magnetic flux $\Phi(a)$ in the hole of the
ring is exactly equal to  the fluxoid $\Phi_f$. The term ``flux
focusing'' is  appropriate because
 $A_{\rm eff}$ is always larger than the actual area of the hole
($\pi a^2$), regardless of the value of $\Lambda$; moreover, $A_{\rm
eff}$ is always in the range $\pi a^2 <  A_{\rm eff} < \pi b^2$.

When
$\Lambda=0$, one has  the limits~\cite{1} $A_{\rm eff}/(\pi a^2)
=(8/\pi^2)(b/a)$ for $a \ll b$ and  $A_{\rm eff}/(\pi a^2) =1 +
(1-a/b)$ for $a \rightarrow b$. A good fit valid for all $a$ in
the range $0 < a < b$ is (see Fig.~14)
 \begin{eqnarray}  
  A_{\rm eff}/ (\pi a^2)= 1/\{\tilde a [\,1 +
    (\pi^2/8 - 1)(1 -\tilde a)^2\,]\}
 \end{eqnarray} with $\tilde a = a/b$.
Values of  $A_{\rm eff}$ obtained from this expression deviate by
less than 0.5\% from the $\Lambda = 0$ calculations of
Ref.\ \onlinecite{1}.~\cite{AABB}  In the limit of large
$\Lambda \gg b$ (which may be applicable to mesoscopic rings
\cite{9,19,20}) one has (see Eq.~(46) below)
  \begin{eqnarray}  
  A_{\rm eff}/ (\pi a^2) = (b^2 -a^2) /[\, 2 a^2 \ln(b/a) \,]\,,
  \end{eqnarray} which is already closely approached for $\Lambda/b
\ge 1$ (see Fig.~14).

The condition that $I = I_1 +I_2 = 0$  (or $I_2 = -I_1$) yields
$A_{\rm eff} = ab\beta_I/\alpha_I$, where $\beta_I =
-\mu_0I_2/bB_a$, shown in Fig.\ 12, was computed in Sec.\ IV and
$\alpha_I = \mu_0aI_1/\Phi_f,$ shown in Fig.\ 6, was computed in
Sec.\ III.
Figure 13 shows the profiles of $J(r)$,
$B(r)$, and $A(r)$ for the  case of flux focusing in a ring with
$a/b=0.4$ for $\Lambda/b =0$, 0.03, and 0.1. Note that the
sheet current  $J(r)$
changes sign inside the superconductor and has zero integral,
$I=0$. As in Figs.~1 and 7, when $\Lambda=0$, $J(r)$ has inverse
square-root infinities at $r=a$ and $r=b$, which were treated in
Ref.\ \onlinecite{1}, but for
$\Lambda >0$, $J(a)$ and $J(b)$ have finite values.

  In Fig.~14 the reciprocal of the effective area $A_{\rm eff}$ is
plotted in the form $\pi a b /A_{\rm eff}$, such that the data for
all values of the penetration depth
$0 \le \Lambda/b \le \infty$ can be presented in one plot. The curve
for $\Lambda =0$ is well fitted by a parabola, Eq.~(29), ranging
from $\pi^2 /8$ at $a/b \to 0$ to 1 at
$a/b \to 1$. The limiting curve for $\Lambda/b \gg 1$, Eq.~(30), is
practically reached already when $\Lambda/b$ exceeds unity. The main
message from this plot is that $A_{\rm eff}$ {\it increases} with
increasing $\Lambda$ for any given $a/b$, and that for
$\Lambda \ne 0$, $A_{\rm eff}/\pi a b$ diverges when $a/b \to 0$,
while for $\Lambda = 0$ it tends to the finite value
$8/\pi^2$. For $\Lambda/b < 0.03$,
$A_{\rm eff}/\pi a b$ has a minimum (since $\pi a b /A_{\rm eff}$
has a maximum) as a function of $a/b$.

   Figures 15, 16, and 17 show the magnetic flux $\Phi(a)$ in the
hole of the ring plotted as $\Phi(a) / \pi a b B_a$, the magnetic
field $B(0)$ in the center of the ring plotted as
$[B(0)/B_a](a/b)$, and the magnetic moment $m/m_0$ of the ring for
the flux-focusing case $I=0$. All three quantities decrease with
increasing $\Lambda/b$.

\section{Ring energies} 

The energy of a superconducting ring with $\Lambda \gg b$ was
calculated in Ref.\ \onlinecite{20} as a function of the fluxoid
number $N$ and the applied magnetic induction $B_a$.
Using the results of the previous sections, we are able to
calculate the energy of a thin ring ($d < \lambda$) with inner and
outer radii
$a$ and $b$  for {\it any} value of $\Lambda$.
The authors of   Ref.\ \onlinecite{20} also calculated,
assuming  $\Lambda \gg b$, the energy
barriers between states $N$ and $N \pm 1$ associated with the
energy cost of moving a vortex or antivortex between the inner and
outer radii.  However, to extend such calculations to the case of
arbitrary
$\Lambda$ is beyond the scope of our paper.

We begin by calculating the total electromagnetic energy of the
ring-coil system, where the coil produces a perpendicular magnetic
induction
$B_a$  at the ring,
  \begin{equation} 
  E=\int\! d^3r \, {\bf B}^2/2\mu_0
  +\int\! d^3r \,\mu_0 \lambda^2 {\bf j}^2/2 .
  \end{equation}
The hole contains the fluxoid
$\Phi_f = N \phi_0$, and  ${\bf B} = {\bf B}_a + {\bf B}_J$, as
discussed in Sec. II. The first integral, which extends over all
space, is the total magnetic-field energy, and the second
integral, which extends only over the volume of the ring, is the
total kinetic energy of the supercurrents.
Using the vector potential, the divergence theorem, and Eqs.\ (2)
and (4), we obtain
  \begin{equation} 
  E=E_a + mB_a/2 + N \phi_0 I/2 ,
  \end{equation}
where $ E_a=\int\! d^3r \, {\bf B}_a^2/2\mu_0$ is the applied
field's magnetic energy in the absence of the ring,
$m = m_1 + m_2$ is the total magnetic moment of the ring, and
$I = I_1 + I_2$ is the total current around the ring.
When the coil currents are controlled so as to maintain
the magnetic induction $B_a$ applied to the ring, the relevant
energy is the Gibbs free energy, which we define as
$G_N(B_a) = E_a + g_N(B_a) = E - W$,
where $W = mB_a$ is the work done by the power supply in
bringing up the magnetic moment to its final value $m$. Thus
  \begin{equation} 
  g_N(B_a) = - mB_a/2 + N \phi_0 I/2 .
  \end{equation}
Eliminating
$m$ and $I$ in favor of the  quantities we have defined and
calculated in Secs. II-V ($m_1 = bN\phi_0 \alpha_m/\mu_0$,
$m_2 = -(8/3)b^3H_a\beta_m$, $I_1 = N\phi_0\alpha_I/\mu_0a$,
$I_2=-bH_a\beta_I$,  $L=\mu_0a/\alpha_I$, and
$\beta_I = \alpha_m)$,  we may express $g_N(B_a)$ in a form
equivalent to that given in  Ref.\ \onlinecite{20} for the case
in which there are no vortices in the annular region $a < r < b$:
  \begin{equation} 
  g_N(B_a) = \epsilon_0[(h-N)^2 + \gamma h^2].
  \end{equation}

The characteristic energy is $\epsilon_0 =\phi_0^2/2L =
\phi_0^2 \alpha_I/2\mu_0 a$. As can be seen most
clearly from Fig.\ 2, for given values of $a$ and $b$,
$\epsilon_0$ decreases monotonically as $\Lambda$ increases
from zero to $\infty$ (see Sec.\ VII); accordingly, $\epsilon_0$
is a monotonically decreasing function of temperature.

The reduced field in Eq.\ (34) is
$ h = B_a/B_0$, where  the scaling field is $B_0 =
\phi_0/A_{\rm eff}$, with $A_{\rm eff}= ab\beta_I/\alpha_I$.
Since $B_0$ is proportional to the
reciprocal of  $A_{\rm eff}$, shown in Fig.\ 14, we see that
$B_0$ decreases from its largest value when
$\Lambda = 0$ to its smallest value when $\Lambda = \infty.$
While $B_0$ is a monotonically decreasing function of temperature,
the range of values spanned by $B_0$ is large only for small $a/b$;
in a narrow ring for which $a \to b$, we find
$B_0 \to \phi_0/\pi b^2$, independent of $\Lambda$.

The constant $\gamma$ in the second term of Eq.\ (34), obtained as
$\gamma = \chi -1 = 8b\beta_m\alpha_I/3a\alpha_m^2 -1,$ is shown in
Fig.\ 18.
%
 \begin{figure}  
\includegraphics[scale=.48]{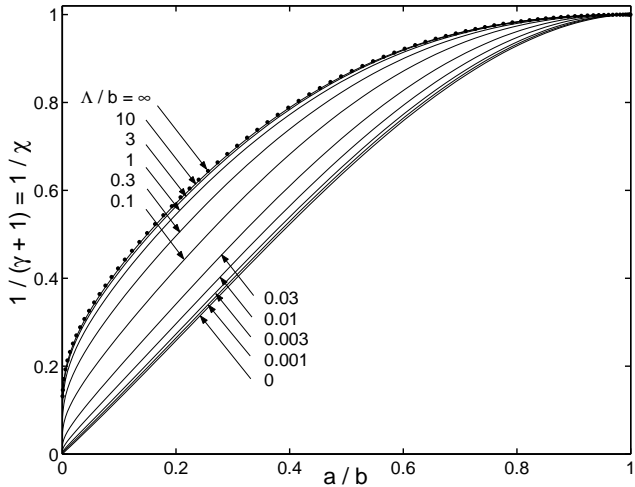}
\caption{\label{fig18} Coefficient $\gamma$ of $h^2= (B_a/B_0)^2$ in
the $N$-independent part of the energy of a superconducting ring in
an applied field [see Eq.\ (34)], plotted as $1/(\gamma +1) = 1/\chi$
to show the entire range of $0 \le a/b \le 1$. The dots mark the
limit $\Lambda \to \infty$, Eq.~(51).
 }
 \end{figure}    
This $N$-independent quadratic term is relatively unimportant,
however, because it is only the first term on the right-hand side of
Eq.\ (34) that determines which quantum state $N$ has
the lowest energy.  Figure 19  exhibits a plot of
$(h-N)^2$ vs $h = B_a/B_0$  for several values of $N$.  From this
plot we can see clearly that the state $N$ has the lowest energy
for values of $h$ in the range $N-1/2 < h < N + 1/2$.
When $h = N - 1/2$, the states $N$ and $N-1$ have the same
energy; similarly, when $h = N + 1/2$, the states $N$ and $N+1$ have
the same energy.
 \begin{figure}  
\includegraphics[scale=.48]{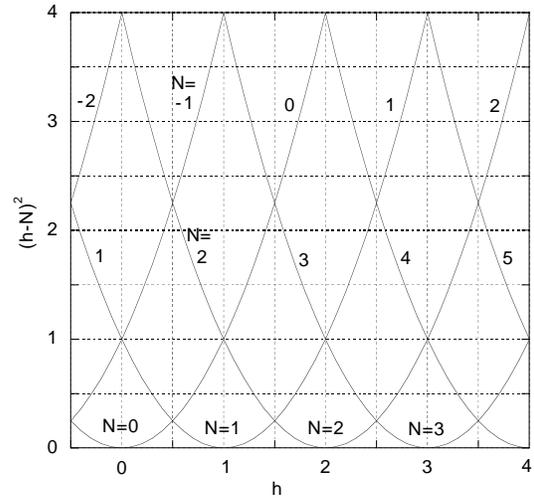}
\caption{\label{fig19} $N$-dependent part of the energy of a
superconducting ring in an applied field [see Eq.\ (34)], plotted as
$(h-N)^2$ vs the reduced field $h = B_a/B_0$.
 }
 \end{figure}    

The temperature dependence of $B_0 = \phi_0/A_{\rm eff}$
leads to the possibility that the energetically favored value of
$N$ may be higher close to $T_c$, where $\Lambda$ diverges, than at
lower temperatures, where $\Lambda$ is much smaller.  This effect,
which may have important consequences for experiments using small
superconducting rings, is greatest  when $a/b$ is small.
For example, suppose that $a = 0.1b$.  At high temperatures for
which $\Lambda/b > 1$, we see from Fig.\ 14 that
$A_{\rm eff} \approx 0.2 \pi b^2$, such that
$B_0 \approx 5 \phi_0 /\pi b^2$. In an applied field of
$B_a \approx 3 \phi_0 / \pi b^2$, $h \approx 0.6$, and the
energetically favored state is $N= 1$.  If at  low temperatures
$\Lambda/b < 0.01$, then $A_{\rm eff} \approx 0.1 \pi b^2$,
about half that at high temperatures.  Since the corresponding
value of $B_0$ is then about twice that at high temperatures,
$h \approx 0.3$, and the quantum number for the state with the
lowest energy becomes $N = 0$.

We emphasize that Eq.\ (34) gives the energy of a ring with
fluxoid number $N$ in the hole when there are {\it no} vortices
in the annular region $a < r < b$. It also states that the
energetically favored value of $N$ is approximately equal to
$B_a/B_0$ and hence increases as $B_0$ increases. However, since
the intervortex spacing for an infinite thin film in an applied
magnetic induction
$B_a$ is of the order of $(\phi_0/B_a)^{1/2}$, we expect that
it will first become energetically favorable for a vortex to sit in
the middle of the annular region, rather than to enter the hole and
increase the value of $N$, when $B_a = B_{c1} \approx \phi_0/w^2$,
where $w = b-a$.  As pointed out in  Ref.\ \onlinecite{20}, for
narrow rings [$w \ll (b-a)$] when $\Lambda/b \gg 1$,
$B_{c1} \approx (2\phi_0/\pi w^2) \ln (2w/\pi \xi)$,  where
$\xi$ is the coherence length or vortex-core radius.

\section{Limit of large penetration depth}  

In the limit $\Lambda/b \gg 1$, which can be realized in small
mesoscopic rings,\cite{9,19,20} analytic expressions for many of
the quantities of interest can be obtained from perturbation
theory.
To lowest order in the small parameter $b/\Lambda$, we may neglect
the current-generated contribution
$A_J$ to the vector potential $A$ in the London equation.
From Eqs.\ (1) and (2) we thus obtain the sheet-current density
$J$:
  \begin{equation} 
  J(r)  = \frac{1}{\mu_0
  \Lambda}\Big(\frac{\Phi_f}{2\pi r} -\frac{1}{2}rB_a\Big).
  \end{equation}
From Eqs.\ (4) and the Biot-Savart law we then obtain the
magnetic moment $m$, total current $I$, and magnetic induction
at the center of the hole $b(0)$ generated by this current:
  \begin{equation} 
  m  = \frac{\pi}{\mu_0
  \Lambda}\Big[\frac{\Phi_f}{4\pi}(b^2-a^2)
  -\frac{B_a}{8}(b^4-a^4)\Big],
  \end{equation}
  \begin{equation} 
  I  = \frac{1}{\mu_0
  \Lambda}\Big[\frac{\Phi_f}{2\pi}\ln(\frac{b}{a})
  -\frac{B_a}{4}(b^2-a^2)\Big],
  \end{equation}
  \begin{equation} 
  b(0)  = \frac{1}{2 \Lambda}
  \Big[\frac{\Phi_f}{2\pi}\frac{(b-a)}{ab}
  -\frac{B_a}{2}(b-a)\Big],
  \end{equation}

For  the case $B_a = 0$ and $\Phi_f > 0$, the self-inductance
$L$ is obtained from Eqs.~(24) and (37).  Expressing $L$ in
terms of the same ratio as shown in Fig.~2, we obtain
  \begin{equation} 
  \frac{L}{\mu_0 a}   = \Big(\frac{\Lambda}{b} \Big)
  \frac{ 2 \pi }{(a/b) \ln(b/a)}.
  \end{equation}
The inverse self-inductance, expressed as the dimensionless
ratio shown in Fig.~6, is
  \begin{equation} 
  \alpha_I =
  \frac {\mu_0 I_1 a}{\Phi_f} = \frac{\mu_0 a}{L}   =
  \frac{1}{2\pi} \Big(\frac{b}{\Lambda}\Big)
  \Big(\frac{a}{b}\Big) \ln \Big(\frac{b}{a}\Big).
  \end{equation}
Expressing the magnetic field $B_1(0)$ at the center of the ring in
terms of the same ratio  as plotted in Fig.~4, we obtain
  \begin{equation} 
  \frac{2 \pi a^2 B_1(0)}{\Phi_f}  =
  \frac{1}{2} \Big(\frac{b}{\Lambda}\Big)
  \Big(\frac{a}{b}\Big) \Big(1-\frac{a}{b}\Big).
  \end{equation}
The magnetic moment $m_1$  expressed as the same
dimensionless ratio plotted in Fig.~5, is
  \begin{equation} 
  \alpha_m =
  \frac{\mu_0 m_1}{b\Phi_f}   =  \frac{1}{4}
  \Big(\frac{b}{\Lambda}\Big)
  \Big[1-\Big(\frac{a}{b}\Big)^2\Big].
  \end{equation}

For the zero-fluxoid case, we obtain the magnetic induction
$B_2(0)$ at the center of the ring by setting $\Phi_f$ = 0 and
adding $B_a$ to Eq.~(38).  Expressing the result in terms of
the same ratio as plotted in Fig.~9, we obtain
  \begin{equation} 
  \frac{B_2(0)}{B_a}   = 1- \frac{1}{4}
  \Big(\frac{b}{\Lambda}\Big)
  \Big(1-\frac{a}{b}\Big).
  \end{equation}
The corresponding magnetic moment is obtained from Eq.~(36) with
$\Phi_f$ = 0.  Expressing the result in terms of the same
dimensionless ratio as plotted in Fig.\ 11, we obtain
  \begin{equation} 
  \beta_m =
  \frac{m_2}{m_0}   =  \frac{3 \pi}{64} \Big(\frac{b}{\Lambda}\Big)
  \Big[1-\Big(\frac{a}{b}\Big)^4\Big].
  \end{equation}
The total current in the ring is obtained from Eq.~(37), again
with $\Phi_f = 0$.  Expressing the result in terms of the same
dimensionless ratio as plotted in Fig.~12, we obtain
  \begin{equation} 
  \beta_I=
  -\frac{I_2}{b H_a}  = \frac{1}{4} \Big(\frac{b}{\Lambda}\Big)
  \Big[1-\Big(\frac{a}{b}\Big)^2\Big].
  \end{equation}
This result is the same as that in Eq.~(42), as expected from the
identity $\alpha_m = \beta_I$, proved earlier.

For the case of flux focusing, we obtain the effective area
$A_{\rm eff} = \Phi_f/B_a$ by setting $I = 0$ in Eq.~(37).
The result is
  \begin{eqnarray}  
  A_{\rm eff} = ab\frac{\beta_I}{\alpha_I} =
  \frac{\pi (b^2 -a^2)}{2 \ln(b/a)}.
  \end{eqnarray}
Expressing $A_{\rm eff}$ in terms of the same ratio as plotted in
Fig.~14, we obtain
  \begin{equation} 
  \frac{\pi a b}{A_{\rm eff}}  =  \pi \frac{\alpha_I}{\beta_I} =
  \frac{2 (a/b) \ln(b/a)}{1-(a/b)^2}.
  \end{equation}
The corresponding magnetic induction at the center of the ring
$B(0)$ is obtained by adding $b(0)$ [Eq.~(38)] to $B_a$ but using
Eq.~(37) and $I=0$ to eliminate $\Phi_f$.  Expressing the result
in terms of the same ratio as plotted in Fig.~16, we obtain
  \begin{equation} 
  \frac{B(0)}{B_a} \frac{a}{b} =  \frac{a}{b}
  +\frac{1}{4} \Big(\frac{b}{\Lambda}\Big) \Big(1- \frac{a}{b}\Big)
  \Big[\frac{(1-a^2/b^2)}{2\ln(b/a)} -\frac{a}{b} \Big].
  \end{equation}
The corresponding magnetic moment $m$ is obtained from
Eq.~(36), using Eq.~(37) and $I=0$ to eliminate $\Phi_f$.
Expressing the result in terms of the same ratio as
plotted in Fig.~17, we obtain
  \begin{equation} 
  \frac{m}{m_0} =
  \frac{3 \pi}{64} \Big(\frac{b}{\Lambda}\Big)
  \Big[1 - \Big(\frac{a}{b}\Big)^2\Big]
  \Big[1 + \Big(\frac{a}{b}\Big)^2 -\frac{(1-a^2/b^2)}{\ln(b/a)}
  \Big].
  \end{equation}

Finally, we calculate the quantities of interest for ring energies,
discussed in Sec.\ VI. When $\Lambda/b \gg 1,$ the characteristic
energy, $\epsilon_0 =\phi_0^2/2L = \phi_0^2 \alpha_I/2\mu_0 a$,
becomes [see Eq.~(40)]
  \begin{equation}  
  \epsilon_0 = \frac{\phi_0^2 \ln(b/a)}{4\pi\mu_0\Lambda},
  \end{equation}
which agrees with that found in Ref.\ \onlinecite{20}, allowing
for the different system of units and definition of
$\Lambda$ used there. For large $\Lambda$, the quantity
$\gamma = \chi -1 = 8b\beta_m\alpha_I/3a\alpha_m^2 -1$ appearing
in Eq.~(34) becomes
  \begin{equation}  
  \gamma = \frac{1+(a/b)^2}{1-(a/b)^2}\ln(b/a) - 1,
  \end{equation}
in agreement with  Ref.\ \onlinecite{20}.

Although the results presented in this section should apply
only when $\Lambda/b \gg 1$, comparison of the above expressions
with the numerical results show that they provide good
approximations already when $\Lambda/b = 1$ and yield excellent
results when  $\Lambda/b \ge 3$.

\section{Summary}  
In this paper we presented a straightforward matrix-inversion
method for the solution of the sheet current, vector potential,
and magnetic field generated by a thin-film ($d < \lambda/2$)
superconducting ring (inner and outer radii $a$ and $b$)
containing a trapped fluxoid
$\Phi_f$ in a perpendicular applied magnetic induction
$B_a$ for values of $\Lambda = \lambda^2/d$ ranging from zero to
infinity. We  used this method to calculate magnetic-field,
current-density, and  vector-potential profiles and numerous
related physical quantities for three important cases:
(i) $\Phi_f > 0$ but $B_a = 0$ (trapped fluxoid, index $n=1$),
(ii) $B_a > 0$ but $\Phi_f = 0$ (zero-fluxoid state, index $n=2$),
     and
(iii) $\Phi_f > 0$ and $B_a > 0$  but no net current around
the ring (flux focusing). We also calculated the Gibbs free
energy of the ring as a function of the quantum number $N$,
where $N = \Phi_f/\phi_0$, and the applied magnetic induction
$B_a$ when no vortices are present in the annular region
between $a$ and $b$.


\acknowledgments
This work was
supported in part by the German Israeli Research Grant Agreement
(GIF) No G-705-50.14/01 and in part by Iowa State University of
Science and Technology under Contract No.\ W-7405-ENG-82 with
the U.S.\ Department of Energy.


\begin{thebibliography}{24}
\bibitem{15} The constancy of the London fluxoid
     $2\pi r [A_J(r)+\mu_0 \Lambda J(r)]$  is confirmed
     by our matrix-inversion method (written in MATLAB) with high
     precision, to 15 significant digits for all $a \le r \le b$.
\end{thebibliography}
\end{document}